# Super-Resolution Based on Deep Operator Networks

Siyuan Yang

## Abstract

We use Deep Operator Networks (DeepONets) to perform super-resolution reconstruction of the solutions of two types of partial differential equations and compare the model predictions with the results obtained using conventional interpolation methods to verify the advantages of DeepONets. We employ two pooling methods to downsample the origin data and conduct super-resolution reconstruction under three different resolutions of input images. The results show that the DeepONet model can predict high-frequency oscillations and small-scale structures from low-resolution inputs very well. For the two-dimensional problem, we introduce convolutional layers to extract information from input images at a lower cost than purer MLPs. We adjust the size of the training set and observe the variation of prediction errors. In both one-dimensional and two-dimensional cases, the super-resolution reconstruction using the DeepONet model demonstrates much more accurate prediction results than cubic spline interpolation, highlighting the superiority of operator learning methods in handling such problems compared to traditional interpolation techniques.

## Introduction

In recent years, artificial intelligence, particularly machine learning based on deep neural networks, has experienced rapid advancements, achieving significant breakthroughs in many areas, such as image recognition and natural language processing. The tremendous potential exhibited by machine learning has made its application to other scientific and engineering challenges (AI4Science) a focal point of considerable interest. Among these challenges, the solution of direct and inverse problems for partial differential equations, the approximation of nonlinear operators , reduced-order modelling of dynamical systems, and the reconstruction of high-dimensional data from low-dimensional representations have attracted widespread attention from researchers.[1][2][3][4] The continual emergence of innovative methods such as Physics-Informed Neural Networks (PINNs), Deep Operator Networks (DeepONets), and Fourier Neural Operators (FNOs) has rendered AI4Science a vibrant and rapidly evolving field.

One of the widely concerned topics within AI4Science is super-resolution, which involves reconstructing high-resolution graphic data from low-resolution graphic data based on machine learning models. In computational fluid dynamics, only grids with sufficiently high spatial density can resolve the small-scale turbulence structures, which often entails significant computational resource consumption. However, modern engineering practices frequently require high-resolution flow simulations to achieve high simulation fidelity, making the reduction of this process's computational cost critically important. With the ability to infer high-dimensional data from low-dimensional data, super-resolution is the proper technique to lower the expense caused by large amounts of grids.

To reconstruct high-resolution images from low-resolution inputs is a common task in computer vision. Without deep learning, researchers have developed various traditional techniques based on interpolation. While these methods require minimal computational resources, they often struggle to handle small-scale structures effectively.[5] When the original images include high-frequency oscillation, interpolation methods often yield results with unacceptable errors. Recently numerous machine-learning-based methods have been introduced to address this limitation. Dong *et al.* applied Convolutional Neural Networks (CNNs), which had demonstrated robust performance in image recognition tasks, to super-resolution, proposing the SRCNN model.[6] Xie *et al.* developed the tempoGAN model, leveraging Multilayer Perceptrons (MLPs) and CNNs to achieve super-resolution for smoke dynamics.[7] However, Generative Adversarial Networks (GANs) typically entail high training costs. Fukami *et al.* implemented a downsampled skip-connection/multi-scale (DSC/MS) mechanism to achieve super-resolution for laminar cylinder wakes and two-dimensional isotropic decaying turbulence, demonstrating superior prediction performance compared to common CNNs in certain cases.[8] In addition to CNNs and GANs, some researchers have explored the direct use of MLPs. Erichson *et al.* applied MLPs to achieve super-resolution for laminar cylinder wakes, two-dimensional isotropic decaying turbulence, and sea surface flows.[9] Although the MLPs have the advantage of simple structures, their training costs are usually high. Nair *et al.* incorporated Proper Orthogonal Decomposition (POD) into MLPs, achieving super-resolution for laminar flat-plate flows and reducing the associated training costs.[10] Furthermore, the introduction of PINNs has led researchers to explore the integration of physics-informed loss functions during the training of super-resolution models, resulting in the development of unsupervised and semi-supervised approaches.[11][12]

In the physical space, the primary objective of super-resolution is to achieve more accurate interpolation—obtaining field information at unsampled points within the field based on data from relatively sparse grids. In Fourier spectral space, this corresponds to extrapolation, where the goal is to infer unknown high-frequency information from known low-frequency structure. Mathematically, it can be proved that if the data represents the solution of certain types of partial differential equations, such problems can be transformed into solvable optimisation problems using Gaussian Process Regression (GPR).[13] Solving this optimisation problem yields a continuous solution function, rather than merely discrete solution values at specific points. It suggests that, at least in certain cases, it is feasible to construct a mapping from low-resolution sample points to a continuous solution function. Consequently, at least in some instances, an operator exists that maps the continuous function obtained through interpolation from low-resolution data to the actual solution function. Therefore, determining this operator would effectively accomplish the goal of super-resolution.

This insight implies that adopting operator learning methods in super-resolution is feasible. Deep Operator Networks (DeepONets) and their variants have emerged as prominent methods for nonlinear operator approximation in recent years. Lu *et al.* initially introduced this approach to solve initial value problems for ordinary differential equations, proving its universal approximation theorem under certain conditions.[2] Subsequently, this method has been successfully applied to solving partial differential equations such as Darcy flows.[14] The development of various DeepONet variants has demonstrated the robust capabilities of this framework across a broader range of applications.[15] This fact motivates the application of

DeepONets to approximate the operators required for super-resolution in the present work. It should be noted that, although it is generally challenging to ascertain whether this operator satisfies the conditions necessary for the universal approximation theorem applicable to DeepONets, its feasibility can always be empirically tested.

This paper aims to implement super-resolution reconstruction of data obtained from the coarse sampling of certain partial differential equation solutions within the DeepONet framework and to compare these results with those obtained through interpolation. The remainder of this paper is organized as follows: Section 2 discusses the methodology and network parameter set-up. Section 3 presents the model's prediction performance using the initial value problem of a one-dimensional Korteweg-de Vries-Burgers (KdV-Burgers) equation with random initial conditions and physical properties, and the boundary value problem of a two-dimensional Poisson equation with a random source term. Additionally, we discuss several observations made during the training and testing of DeepONet models. Section 4 provides the conclusion.

## Methodology

For a continuous input function $u$, we can introduce an operator $O(*)$ to map it to another continuous function $v$, which can be expressed as

$$v(y) = O(u)(y) \tag{1}$$

Lu Lu *et al.* pointed out the universal approximation theorem for operator learning, which states that when $u$ and $O(*)$ satisfy certain conditions, DeepONets have the ability to approximate $O(*)$ with arbitrary demanded accuracy, provided there are no restrictions on the depth of the network or the number of neurons.[2] Specifically, if we denote the neural network parameters as $w$, the required operator can be fitted by

$$w = argmin_w \mathcal{L}\big(v(y), \mathcal{F}(u;w)(y)\big) \tag{2}$$

where $\mathcal{L}(*)$ is the loss function, $\mathcal{F}(*)$ is the operator represented by DeepONets, and $v$ is the targeted function in the training set.

In the typical DeepONet framework, there are two neural networks, referred to as the branch net and the trunk net. The branch net processes the sampled input function $u$, while the trunk net handles the positional information $y$. The results from both networks are combined through a dot product to produce the model output $\mathcal{F}(u;w)(y)$. For a given input pair $(u, y)$, if the target function $v$ is scalar, the model will output a value to approximate $v$ at the specified position. If both the input function and the target function are defined on a subset of $\mathcal{R}^1$, both the branch net and the trunk net can be simply implemented using MLPs. It has been validated in the application of solving initial value problems for ordinary differential equations with DeepONets.[2] We will use this method to achieve super-resolution for the 1D KdV-Burgers equation solution and compare the results with interpolation results.

The MLP is the simplest neural network architecture. The hidden layers between the input layer and the output layer consist of fully connected layers, meaning each neuron is connected to all neurons in the previous layer. The output $c_i^{(l)}$ of the $l$-th layer can be expressed using the input

$c_j^{(l-1)}$ from the previous layer, the activation function $\phi$, the weights $\omega$, and the biases $b$.

$$c_i^{(l)} = \phi\left(\sum_j \omega_{ij}^{(l)} c_j^{(l-1)} + b_i^{(l)}\right) \tag{3}.$$

Using backpropagation to minimise the loss function, the weights and biases are optimised, enabling the model to fit nonlinear relationships.

Since the hidden layers in MLPs are fully connected, with each neuron receiving information from all neurons in the previous layer, MLPs should be able to process global information. However, when the input function is defined on a subset of $\mathcal{R}^2$ or even higher-dimensional spaces, the significant increase in degrees of freedom leads to a dramatic rise in the number of parameters in the fully connected layers. In this case, using MLPs becomes too costly. In contrast, CNNs can reduce the amounts of the model's parameters due to their local connectivity and weight-sharing mechanisms, while still effectively extracting features. CNNs have already demonstrated strong capabilities in handling high-dimensional image data, making the introduction of CNNs into the branch net a reasonable choice. Therefore, for the 2D case, we incorporate two convolutional layers into the branch net, while the trunk net remains a simple MLP. Similar approaches have been employed in the application of DeepONets to solve Darcy flows.[14]

In this work, we take low-resolution data as the input to the branch net, which can be understood as using the continuous function obtained by interpolating the low-resolution data as the input function $u$, with high-resolution data as the reference of target functions. During training and prediction, the input to the trunk net consists of the coordinates of sample points of the high-resolution data. Specifically, for the 2D case, in the branch net, we first replicate the low-resolution data across 200 channels, then apply a convolutional layer with a kernel size of 1 and a stride of 1, setting the output channel count to 100. We then perform Batch Normalisation and use ReLU (Rectified Linear Units) as the activation layer. Next, the output data with 100 channels is passed through a second convolutional layer with a kernel size of 2 and a stride of 1, setting the output channel count to 40, followed by Batch Normalisation and ReLU activation. After flattening the resulting data, we feed it into an MLP with four fully connected layers, incorporating Batch Normalisation and Softplus as the activation function, to extract the information from the low-resolution data.

The network structure used in this work is shown in Figure 1. Here, we choose 2D Poisson equations with random source terms as an example, and the details of this case will be discussed in the next section. The dataset used for training and testing was generated using numerical simulation via the spectral method. The low-resolution data used as input was obtained by pooling the numerical computation results.

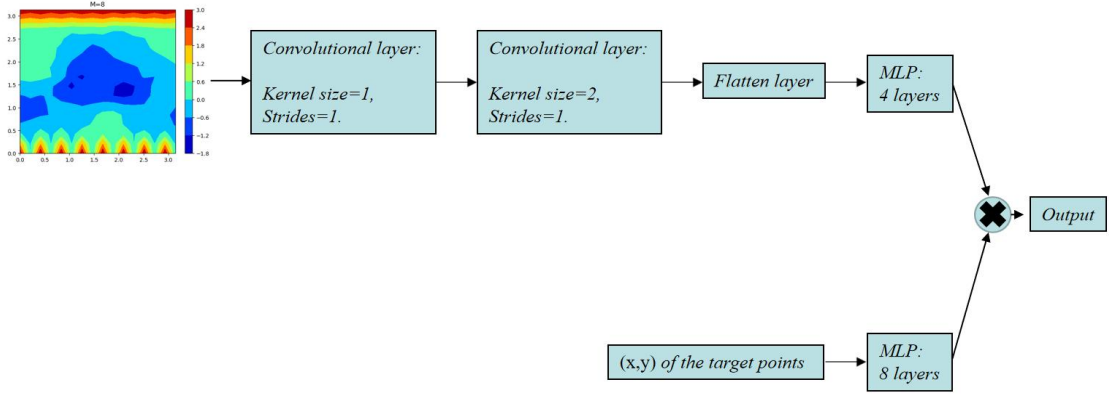

Figure 1. The schematic of the DeepONet structure in this work. This figure takes the case of using low-resolution 2D Poisson equation solutions as input as the example.

Pooling is a common downsampling operation used in convolutional neural networks. Following the work of Fukami *et al.*, we chose two methods: average pooling and max pooling, and we will demonstrate the prediction performance of the machine learning model under different pooling methods.[8] Taking the 2D case as an example, average pooling can be defined as

$$v_{ij}^{LR} = \frac{1}{M^2} \sum_{s,t \in P_{i,j}} v_{st}^{HR} \qquad (4),$$

while max pooling can be defined as

$$v_{ij}^{LR} = max_{s,t \in P_{i,j}} v_{st}^{HR} \qquad (5).$$

We can adjust $M$ to change the resolution of the input. The larger the value of $M$, the lower the resolution of the image after pooling. Taking the 2D case as an example, we will use $M$ values of 4 (medium resolution), 8 (low resolution), and 16 (super-low resolution) to discuss the impact of $M$ on the model's prediction performance and interpolation results. The low-resolution images generated by the two pooling methods with different $M$ values are shown in Figure 2.

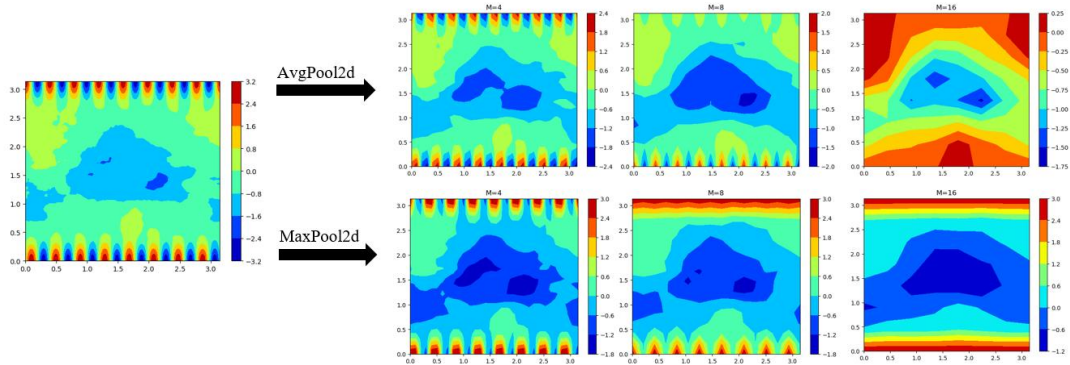

Figure 2. One example of the low-resolution input data after average and max pooling.

In the training process of this work, MSE (Mean Square Error) was uniformly selected as the loss function, and the adaptive moment estimation (Adam) algorithm (Kingma *et al.*) was chosen as the optimiser to learn the network parameters.[16]

In the following sections, we will present the super-resolution results for the solution of the 1D KdV-Burgers equation with random initial conditions and physical parameters, as well as the solution of the 2D Poisson equation with a random source term, using both DeepONet models without convolutional layers and those with convolutional layers. This will verify the feasibility of using DeepONets for super-resolution and compare the results with interpolation-based predictions.

## Results

### 1. 1-dimensional KdV-Burgers Equation

We first use a one-dimensional problem to verify the ability of the DeepONet model to perform super-resolution tasks. We obtain data by numerically simulating the initial value problem of the one-dimensional KdV-Burgers equation under periodic boundary conditions using the open-source software DeDalus3.[17] The KdV-Burgers equation is a hybrid of the KdV equation and the Burgers equation. The KdV equation is commonly used to describe shallow water waves and has soliton solutions, while the Burgers equation is the simplest equation that can describe shock waves. They are basic model equations in computational fluid dynamics, often used to validate the effectiveness of numerical schemes. The specific hybrid form used here is

$$\frac{\partial u}{\partial t} + u\frac{\partial u}{\partial x} - b\frac{\partial^3 u}{\partial x^3} = a\frac{\partial^2 u}{\partial x^2} \tag{6},$$

where $a$ is a random number uniformly distributed over $[10^{-4}, 6 \times 10^{-4}]$, and $b$ is a random number uniformly distributed over $[1.5 \times 10^{-4}, 2.5 \times 10^{-4}]$. The computational domain for the numerical simulation is [0,10], with a uniform grid consisting of 1024 points. The initial condition is given by

$$u(x, 0) = \frac{1}{2n} ln\left(1 + \frac{ch^2 n}{ch^2 n(x-2)}\right) \tag{7},$$

where n is a random integer uniformly distributed over $[-90,110]$.

The solutions of the KdV-Burgers equation obtained under these parameter settings exhibit typical characteristics of both equations, featuring large gradient regions and propagating high-frequency oscillations. It is difficult to infer such structures accurately using traditional interpolation methods, so these solutions can be used to validate the ability of the DeepONet model to perform super-resolution tasks.

To demonstrate the capability of the DeepONet model in handling one-dimensional super-resolution tasks, we use the interp1d function from SciPy for cubic spline interpolation and compare the interpolation results with the model predictions. We perform max pooling and average pooling on the original data to obtain low-resolution data as input for the model. 900 pairs of pooled and original data are selected as the training set. During max and average pooling, we take $M = 8$ to represent medium-resolution input, $M = 16$ as low-resolution input, and $M = 32$ as super-low-resolution input, corresponding to resolutions of 128 pixels, 64 pixels, and 32 pixels, respectively. In Figure 3, the top image shows the original data before pooling, with the left side showing the results under max pooling and the right side showing the results under average pooling. The first row represents the input data, the second row shows the results predicted using interpolation, and the third row shows the results obtained by the DeepONet model.

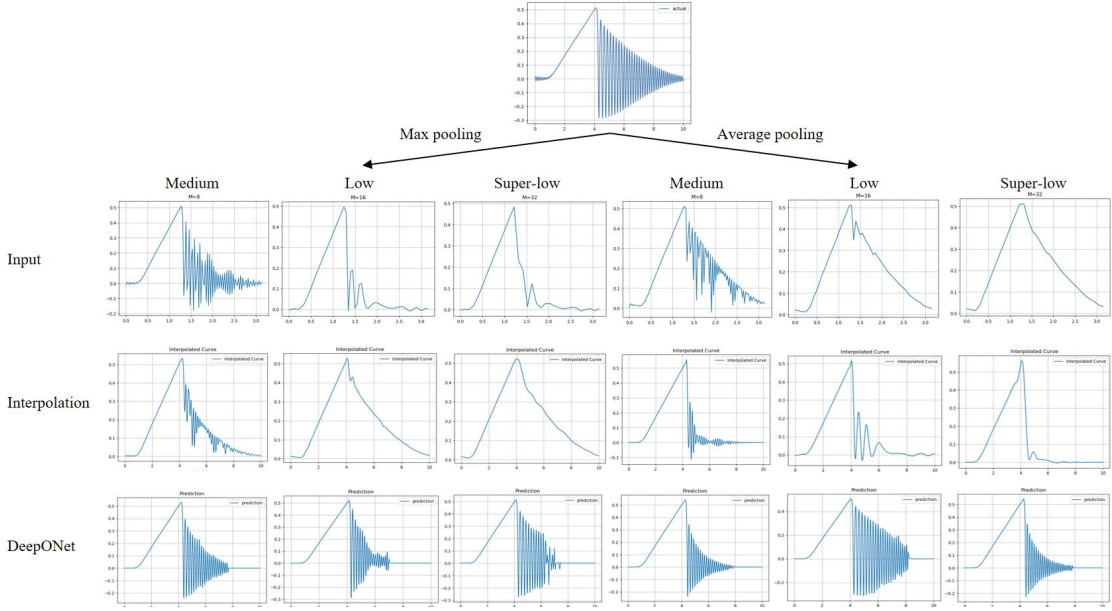

Figure 3. One example in the test set to show the super-resolution reconstruction performance of interpolation and DeepONet models for the solution of 1D KdV-Burgers equation

We summarize the L2 errors of the prediction shown in Figure 3 obtained using different pooling methods and different values of M in Table 1. The L2 error is defined as

$$\epsilon = \frac{\|x^{HR} - \mathcal{F}(x^{HR})\|_2}{\|x^{HR}\|_2} \tag{8}.$$

| *Max Pooling* | Medium $M = 8$ | Low $M = 16$ | Super-low $M = 32$ |
|---|---|---|---|
| *Interpolation* $\epsilon$ | 76016.9 | 80388.2 | 85479.2 |
| *DeepONet* $\epsilon$ | 90.8689 | 135.2939 | 160.927 |

Table 1a L2 error for inputs obtained by Max Pooling in Figure 3

| *Average Pooling* | Medium $M = 8$ | Low $M = 16$ | Super-low $M = 32$ |
|---|---|---|---|
| *Interpolation* $\epsilon$ | 72009.1 | 71113.8 | 71496.1 |
| *DeepONet* $\epsilon$ | 94.2727 | 69.1431 | 115.017 |

Table 1b L2 error for inputs obtained by Average Pooling in Figure 3

Let us analyse the results obtained through interpolation first. As shown in Table 1, the interpolation method performs better with average pooling than with max pooling. The error obtained through average pooling is smaller than that obtained through max pooling across different values of $M$. Similarly, it can also be observed that for max pooling, the L2 error of cubic interpolation increases as $M$ increases. From Figure 3, it can be seen that cubic interpolation almost fails to predict the oscillations in the solution of the one-dimensional KdV-Burgers equation and cannot accurately predict the shape of the shock wave peaks.

We further examine the super-resolution reconstruction results of the DeepONet model for low-resolution inputs. The table shows that, for both max pooling and average pooling, the L2 errors of the results predicted by the model we used show no significant difference, and the error is significantly smaller than that of the interpolation predictions. For all three different resolutions, the DeepONet model can predict the oscillatory characteristics well, which are not captured by the interpolation method.

We adopted 100 pairs of pooled and original data as the test set and compared the results predicted by the DeepONet model with those predicted by interpolation, using two different pooling methods and three different resolutions, as shown in Figure 4. Since the L2 errors obtained from interpolation and DeepONet predictions exhibit differences in orders of magnitude, we take the logarithm (base 10) of the average L2 error to facilitate graphical representation. It can be seen that the average L2 error of the results predicted by the DeepONet model also increases as the resolution decreases (i.e., as $M$ increases), but the rate of increase is much smaller than that of the interpolation results. Moreover, the L2 error from the super-resolution analysis of the DeepONet model is significantly smaller than that from interpolation predictions. For example, in the case of $M = 16$ (which corresponds to a resolution of 64 pixels), the average L2 error of the input image after max pooling predicted by interpolation is 80388.2, while the L2 error predicted by the DeepONet model is only 135.2939. The difference between the two is two orders of magnitude, demonstrating that the super-resolution predictions of the DeepONets model for the solution of the one-dimensional KdV-Burgers equation are far more accurate than those of interpolation. The lower the resolution of the input image, the more pronounced this advantage becomes.

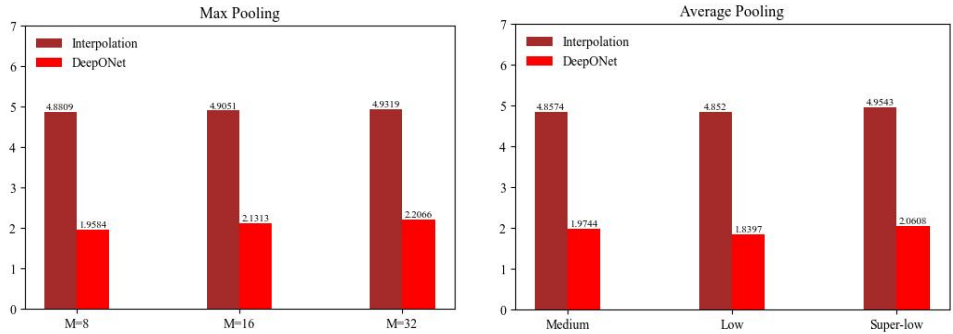

Figure 4. Average L2 error on the test set for 1D KdV-Burgers equation

The training of the DeepONets models for the KdV-Burgers equation is carried out on a single GPU (NVIDIA GeForce RTX 4060 Laptop graphics processing unit). We distribute the data across different units for training. Since the size of data and models here is not very large, there is no need to distribute the training across multiple GPUs, and training across different units has still significantly improved computation speed. During training, with the six different parameter conditions, the number of neurons in the network varies which results in some variation in the training time. The longest training time occurs with the model trained under max pooling at medium resolution, taking 56 minutes and 57.6 seconds, with a computation speed of approximately 2.87 seconds per epoch. It is worth noting that the time spent training one-dimensional data is not very long. Although DeepONet does not have an advantage in prediction efficiency compared to interpolation methods, as mentioned earlier, the super-resolution predictions obtained by the DeepONet model are far more accurate than those from interpolation methods.

Subsequently, we can obtain the prediction results from both interpolation and DeepONet models similarly using different training data sizes. When $M = 16$, using training sets with sizes of 45, 90, 225, 450, and 900, Figure 5 shows the average L2 error on the test sets of the DeepONet models.

The overall trend of the graph shows that as the size of the training dataset increases, the average L2 error of the predictions made by the DeepONet model decreases. In other words, the larger the training dataset, the more accurate the predicted results. However, the graph does not exhibit a strictly monotonic downward trend, which is supposed to stem from the randomness in data selection.

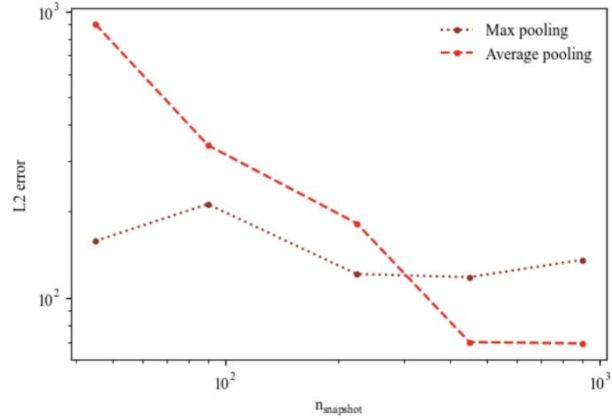

Figure 5. the relationship between the average L2 error on the test set and the training set size for the 1D KdV-Burgers equation

## 2. 2-dimensional Poisson Equation

To further demonstrate the super-resolution capability of the DeepONet model, we consider the boundary value problem of a two-dimensional Poisson equation with a random source term. We still use the open-source software Dedalus3 for numerical simulations employing spectral methods to obtain the data.[17] The governing equation used in the simulations is

$$\nabla^2 u = f \qquad (9),$$

where $f$ is a random number in the range [0,100] at each sampling point. The computational domain is $[0,\pi]\times[0,\pi]$, and the grid resolution used for the numerical simulations is 128×128. The boundary conditions are

$$u(0,y) = u(\pi,y), y \in [0,\pi] \qquad (10.a),$$
$$u(x,0) = 3\sin(16x), x \in [0,\pi] \qquad (10.b),$$
$$u(x,\pi) = 3\cos(16x), x \in [0,\pi] \qquad (10.c).$$

That is, periodic boundary conditions are applied in the vertical direction, while the values of the solution are specified on the horizontal boundaries. The oscillatory boundary conditions, as in (10. b) and (10. c), ensure that the solution contains small-scale structures to evaluate DeepONet models' ability to predict small-scale structures that are hard to address with interpolation. Although the Poisson equation is simple in form, its solution is fundamental to many classic numerical methods for solving Navier-Stokes equations. In numerical simulations of incompressible fluids, how to quickly and accurately solve the pressure Poisson equation

$$\nabla^2 p = -\rho \nabla \cdot (v \cdot \nabla v) \tag{11}$$

has always been a key focus of researchers. Therefore, using the Poisson equation as an example can not only validate the effectiveness of the DeepONet super-resolution model but also has practical significance.

We apply the DeepONet model described in Section 2, which includes convolutional layers, to the system formed by equation (9) and boundary conditions (10. a, b, c) for 2D super-resolution. For comparison, we still use the cubic spline interpolation method, implemented via the cubic option of the interp2d function in the SciPy library, which is a common method in image processing tasks for reconstructing high-resolution images from low-resolution ones. By performing max and average pooling on the 128×128 data obtained from the simulation, we obtain coarse data as input. After feeding the coarse data and target points' coordinates into the neural network, we can predict the value of $u$ at the target points. We obtain super-resolution results by making predictions at each target point for the same coarse data, which can then be compared with interpolation.

We first show the DeepONet model's performance on the test set when using 900 pairs of pooled and original data as the training set. Here we set $M = 4$ as medium-resolution input, $M = 8$ as low-resolution input, and $M = 16$ as super-low-resolution input, corresponding to 32 pixels×32 pixels, 16 pixels×16 pixels, and 8 pixels×8 pixels, respectively. In Figure 6, the top image is the original data used as a reference. On the left are the results using max-pooled data as input, and on the right are the results using average-pooled data as input. The first row shows the input data, the second row shows the results obtained by interpolation, and the third row shows the results predicted by DeepONet models.

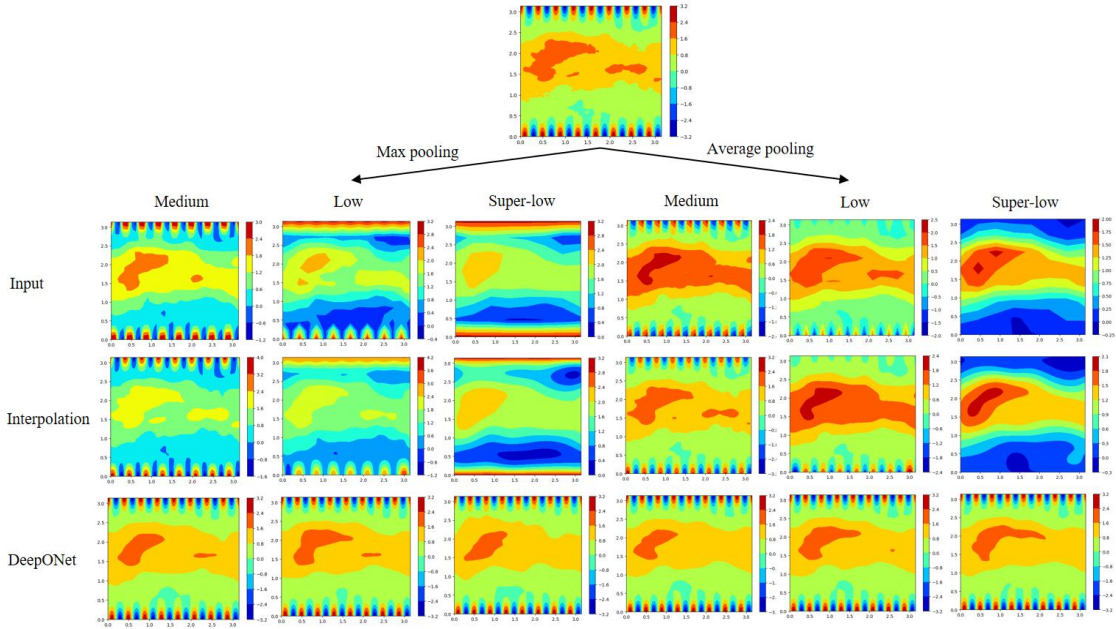

Figure 6. One example in the test set to show the super-resolution reconstruction performance of interpolation and DeepONet models for the solution of 2D Poisson equation

We summarize the L2 errors of the prediction shown in Figure 5 obtained using different pooling methods and M values for each method in Table 2.

| Max Pooling | Medium $M = 4$ | Low $M = 8$ | Super-low $M = 16$ |
|---|---|---|---|
| Interpolation $\epsilon$ | 0.4424 | 0.9333 | 1.0746 |
| DeepONet $\epsilon$ | 0.0821 | 0.0824 | 0.1538 |

Table 2a L2 error for inputs obtained by Max Pooling in Figure 6

| Average Pooling | Medium $M = 4$ | Low $M = 8$ | Super-low $M = 16$ |
|---|---|---|---|
| Interpolation $\epsilon$ | 0.2347 | 0.5957 | 0.6406 |
| DeepONet $\epsilon$ | 0.0886 | 0.0705 | 0.1111 |

Table 2b L2 error for inputs obtained by Average Pooling in Figure 6

We still first examine the predictive performance of the interpolation method. The phenomena are very similar to the 1D results in Section 3.1. It can be observed that the interpolation method performs significantly better for inputs obtained via average pooling compared to those obtained via max pooling. This may be because, for the specific case considered here, average pooling better preserves information about small-scale structures than max pooling. For both pooling methods, the L2 error of interpolation increases as the M value increases, i.e., as the amount of input information decreases. From the images, it can be seen that for medium-resolution inputs, the interpolation method can still reasonably predict the small-scale structures near the boundary caused by oscillations. However, when the input image resolution drops to $M = 8$, interpolation can no longer accurately predict the oscillations near the boundary, and for super-low-resolution inputs, its prediction contains no oscillation information at all. This phenomenon is natural: as long as the structure of interest is smaller than the filtering window used during pooling, the corresponding structural information will not appear in the input image. Since the interpolation method does not contain any prior knowledge specific to the problem class, it is impossible to infer information filtered out by the input image. The ability to infer high-wavenumber information is precisely what the super-resolution problem demands, highlighting the necessity of introducing machine learning methods.

Next, we can examine the performance of DeepONet models. Many observations here are also highly similar to the 1D results. For both pooling methods, the DeepONet model shows no significant difference in L2 error for this test case. For the three input resolutions—medium, low, and super-low—the DeepONet models can accurately predict the small-scale structures near the boundary caused by oscillations, in stark contrast to the interpolation method. The reason for this is that for the specific problem formed by equation (9) and boundary conditions (10. a,b,c), the supervised learning model abstracts and stores information about the properties of the specific problem as parameters during training. This allows for the inference of high-wavenumber information from low-wavenumber inputs, which has been filtered out earlier. Since DeepONets are devised for operator learning, the explanation provided earlier specifically means that the DeepONet model abstracts an operator during training that maps the spectrum defined in the low-wavenumber range to the spectrum defined over a broader range of wavenumbers. The results show that different pooling methods do not significantly impact the fitting process of this operator.

For a test set consisting of 100 data pairs, the average relative errors of the DeepONet model and the cubic interpolation method in super-resolution prediction under different pooling methods and input resolutions are shown in Figure 7. Taking the case of M=8 as an example, for input images obtained via max pooling, the average L2 error of the interpolation method for super-resolution prediction is 1.0354, while for the DeepONet model, it is 0.1076, which is only one-ninth of the former. This indicates a significant improvement in the ability of operator learning methods to predict small-scale structures compared to interpolation methods in 2-dimensional problems. Like the 1D results, this improvement is more pronounced when the resolution of the input image is lower.

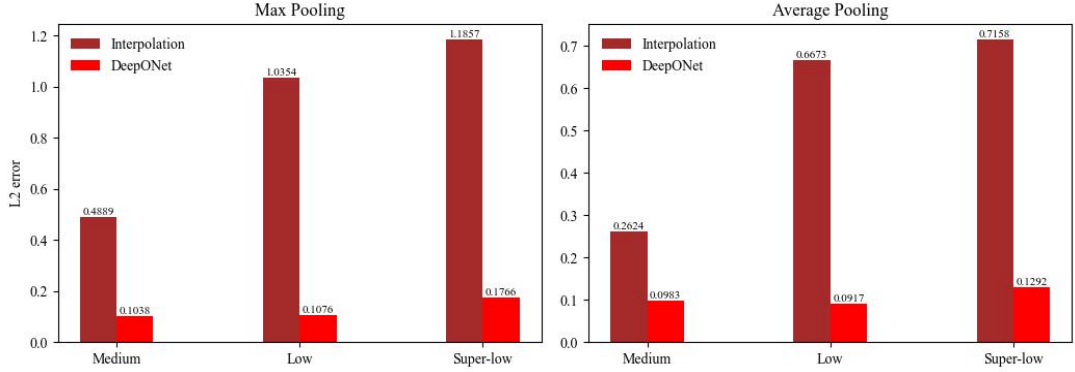

Figure 7. Average L2 error on the test set for 2D Poisson equation

The model training is conducted on a single NVIDIA GeForce RTX 3060 Laptop GPU. Since the number of neurons in the network varies under the six parameter conditions, the training time is different for each model. The longest training time was for the model corresponding to medium input resolution under average pooling, taking about 21.7 seconds per epoch, with a total training time of approximately 7.24 hours. The training time for the model corresponding to low input resolution under max pooling is more representative, taking approximately 9.4 seconds per epoch, with a total training time of 3.21 hours. Although this time cost may seem large, for similar problems—i.e., those determined by the same governing equation and boundary conditions—the DeepONet model only needs to be trained once. This time cost is necessary to fully extract information from the training set to fit the operator and achieve more accurate predictions. It is worth noting that since the DeepONet super-resolution model needs to repeat the same calculations for all target points during prediction, we cannot expect it to be faster than interpolation in serial computation. However, the efficiency should be improvable through parallel computation.

The number of snapshots included in the training set can also be adjusted to examine the impact of data size on the prediction performance of the DeepONet super-resolution model. Figure 8 shows the average L2 error on the test set for models trained on datasets of different sizes with $M = 8$ under two pooling methods. The model's prediction ability improves significantly as the training set size increases, and overall, the model with low-resolution images obtained through average pooling as input achieves smaller prediction errors. Notably, even the DeepONet super-resolution model with the weakest prediction ability, when $n_{snapshot} = 45$, has a lower average L2 error than interpolation (the typical values of the interpolation error are listed in Table 2).

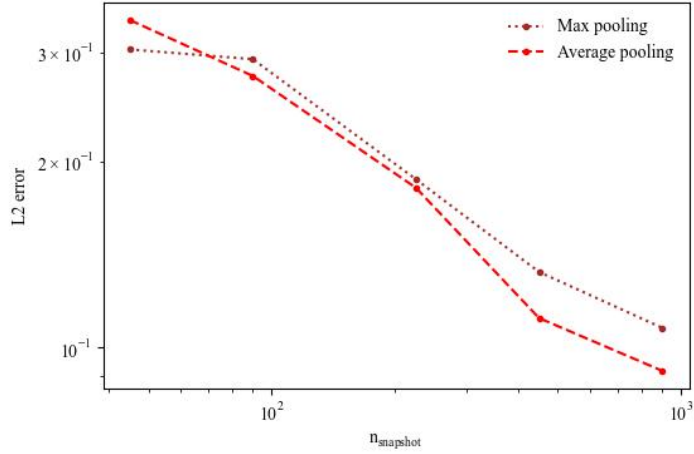

Figure 8. the relationship between the average L2 error on the test set and the training set size for the 2D Poisson equation

## Conclusion

We have utilized the DeepONet framework to perform super-resolution reconstruction of the downsampled numerical solutions of several differential equations. In the model for one-dimensional KDV-Burgers equations, we have constructed the branch net and trunk net using MLP. Additionally, in the model for two-dimensional Poisson equations, we have introduced CNNs, employing two convolutional layers in the branch net while retaining the MLP for the trunk net. The results indicate that, under the parameter conditions we have used, the predictions made by DeepONets in these two types of problems are significantly superior to those obtained using the commonly employed cubic interpolation. Furthermore, this work has employed average and max pooling methods to downsample the high-resolution data and compare the models' prediction performance for the low-resolution data obtained by different pooling methods. During training, MSE has been used as the loss function for optimisation. Similar to other machine learning models, it has been observed that the DeepONet model can learn important features, such as high-frequency oscillations, that interpolation methods fail to capture. This fact highlights the advantage of more degrees of freedom of parameters offered by neural networks over conventional interpolation methods.

The accurate results obtained with DeepONets demonstrate the potential and application prospects of operator learning in super-resolution reconstruction. This also raises a question: can DeepONets and other operator learning techniques be applied to broader super-resolution tasks and other AI4Science problems beyond solving differential equations? With the support of larger datasets, DeepONets, and other operator learning methods should be able to conduct more accurate and extensive super-resolution reconstruction.

## Acknowledgements

Siyuan Yang thanks Yixin Zhang from Peking University for his inspiring discussions.